\documentclass[galaxies,review,accept,pdftex,moreauthors]{Definitions/mdpi} 

\firstpage{1} 
\pubvolume{1}
\issuenum{1}
\articlenumber{0}
\pubyear{2024}
\copyrightyear{2024}
\externaleditor{Academic Editor: Firstname \linebreak Lastname}
\datereceived{29 June 2024} 
\daterevised{31 July 2024} 
\dateaccepted{1 August 2024} 
\datepublished{ } 
\hreflink{https://doi.org/} 

\def\aap{Astron. Astrophys.}                
\def\aj{Astron. J.}

\def\apj{Astrophys. J.}   
\def\apjl{Astrophys. J. Lett.}   

\def\mnras{Mon. Not. R. Astron. Soc.}

\def\apss{Astrophys. Space Sci.}  

\newcommand\degr{\ensuremath{^\circ}}

\usepackage{comment}

\Title{The IXPE View of Neutron Star Low-Mass X-ray Binaries}

\TitleCitation{The IXPE View of Neutron Star Low-Mass X-ray Binaries}

\Author{Francesco Ursini $^{1,}$*\orcidA{}, 
Andrea Gnarini $^{1}$\orcidB{},
Fiamma Capitanio $^{2}$\orcidC{},
Anna Bobrikova $^{3}$\orcidD{},
Massimo Cocchi $^{4}$\orcidE{},
Alessandro Di Marco $^{2}$\orcidF{},
Sergio Fabiani $^{2}$\orcidG{},
Ruben Farinelli $^{5}$\orcidH{},
Fabio La Monaca $^{2,6,7}$\orcidI{},
John Rankin $^{2,8}$\orcidJ{},
Mary~Lynne~Saade $^{9,10}$\orcidK{} 
and Juri Poutanen $^{3}$\orcidL{}
}

\AuthorNames{Francesco Ursini, Andrea Gnarini, Fiamma Capitanio, Anna Bobrikova, Massimo Cocchi, Alessandro Di Marco, Sergio Fabiani, Ruben Farinelli, Fabio La Monaca, John Rankin, M.~Lynne~Saade and Juri Poutanen}

\AuthorCitation{Ursini, F.; Gnarini, A.; Capitanio, F.; Bobrikova, A.; Cocchi, M.; Di Marco, A.; Fabiani, S.; Farinelli, R.; La Monaca, F.; Rankin, J.;~et~al.}

\address{%
$^{1}$ \quad Dipartimento di Matematica e Fisica, Università degli Studi Roma Tre, Via della Vasca Navale 84, \mbox{00146 Roma, Italy;} francesco.ursini@uniroma3.it (F.U.); andrea.gnarini@uniroma3.it (A.G.)\\
$^{2}$ \quad INAF Istituto di Astrofisica e Planetologia Spaziali, Via del Fosso del Cavaliere 100, 00133 Roma, Italy; fiamma.capitanio@inaf.it (F.C.); alessandro.dimarco@inaf.it (A.D.M.); sergio.fabiani@inaf.it (S.F.); fabio.lamonaca@inaf.it (F.L.M.); john.rankin@inaf.it (J.R.)\\
$^{3}$ \quad Department of Physics and Astronomy, University of Turku, FI-20014 Turku, Finland; anna.a.bobrikova@utu.fi (A.B.); juri.poutanen@utu.fi (J.P.)\\
$^{4}$ \quad INAF Osservatorio Astronomico di Cagliari, via della Scienza 5, I-09047 Selargius, Italy; massimo.cocchi@inaf.it (M.C.) \\
$^{5}$ \quad INAF Osservatorio di Astrofisica e Scienza dello Spazio di Bologna, Via P. Gobetti 101, I-40129 Bologna, Italy; ruben.farinelli@inaf.it (R.F.)\\
$^{6}$ \quad Dipartimento di Fisica, Universit\`{a} degli Studi di Roma ``Tor Vergata'', Via della Ricerca Scientifica 1, \mbox{00133 Roma, Italy} \\
$^{7}$ \quad Dipartimento di Fisica, Universit\`{a} degli Studi di Roma ``La Sapienza'', Piazzale Aldo Moro 5, \mbox{00185 Roma, Italy} \\
$^{8}$ \quad INAF Osservatorio Astronomico di Brera, Via E. Bianchi 46, 23807 Merate, Italy \\
$^{9}$ \quad Science \& Technology Institute, Universities Space Research Association, 320 Sparkman Drive, \mbox{Huntsville, AL 35805, USA}; mlsaade@usra.edu (M.L.S.)\\
$^{10}$ \quad NASA Marshall Space Flight Center, Huntsville, AL 35812, USA
}

\corres{Correspondence: francesco.ursini@uniroma3.it}

\abstract{
Low-mass X-ray binaries hosting weakly magnetized neutron stars (NS-LMXBs) are among the brightest sources in the X-ray sky. Since 2021, the Imaging X-ray Polarimetry Explorer (IXPE) has provided new measurements of the X-ray polarization of these sources. IXPE observations have revealed that most NS-LMXBs are significantly polarized in the X-rays, providing unprecedented insight into the geometry of their accretion flow. 
In this review paper, we summarize the first results obtained by IXPE on NS-LMXBs, the emerging trends within each class of sources (atoll/Z), and possible physical interpretations.
}


\begin{document}

\section{Introduction}
Accreting, weakly magnetized NS-LMXBs are very bright X-ray sources and~represent excellent laboratories for studying the physics of accretion on compact objects. 
They are divided into two main classes, Z and atoll, according to the pattern they trace on X-ray color--color diagrams \citep{hasinger&vanderklis}. Atoll sources are less luminous and have lower accretion rates compared with Z sources.
The X-ray spectrum of weakly magnetized NS-LMXBs is generally well described by a soft thermal component and a hard Comptonization component, but~their physical origin is still uncertain. In~principle, the~soft component could be due either to the accretion disk or to the neutron star, while the hard component could originate in a hot corona, in~a boundary layer (BL) between the disk and the NS~\cite{popham&sunyaev2001}, or~a more extended spreading layer (SL) around the NS~\cite{inogamov&sunyaev1999,suleimanov&poutanen2006}. Variability studies showed that the spectrum of the most rapidly variable component is hard, and~is likely associated with the BL rather than the disk~\cite{gilfanov2003,revnivtsev2006}. 
Furthermore, the~primary X-ray emission can be reprocessed by the accretion disk, resulting in a reflection component, which typically consists of a Compton hump at 20--30 keV and fluorescent emission lines (see~\cite{ludlam2024} and the references therein). 

X-ray spectroscopy and timing techniques allow us to study the physical properties of NS-LMXBs in depth (see \citep{disalvo2023} for a recent review), but~their geometry remains elusive. X-ray polarimetry crucially adds two more observables, namely the polarization degree (PD) and angle (PA), which depend on the geometry of the accretion flow. By~measuring X-ray polarization, we can distinguish between radially extended regions such as the accretion disk, and~vertically or spherically extended structures such as the BL or SL \citep{gnarini2022,farinelli2024}. Radiation reflected off the disk is also expected to be significantly polarized \citep{lapidus&sunyaev1985}. 

Before the launch of the Imaging X-ray Polarimetry Explorer (IXPE) \citep{weisskopf2022,soffitta2021}, the~X-ray polarization properties of NS-LMXBs were almost unknown. The~only source with a significant measurement was Sco~X-1, from~a 15-day observation with OSO-8 \citep{long1979} and, more recently, a~322-day observation with PolarLight \citep{long2022}. A~marginal detection of Cyg~X-2 was also obtained by OSO-8 \citep{long1980}. As~of December 2021, IXPE has re-opened the X-ray polarimetric window. Thanks to its greatly improved sensitivity, IXPE has measured the X-ray polarization in the 2--8 keV band of several NS-LMXBs, with~only $\sim$1-day exposures. 
Moreover, IXPE is capable of performing energy- and time-resolved~spectropolarimetry.

This paper summarizes the first X-ray spectro-polarimetric results obtained with IXPE on NS-LMXBs during the first two-year observational campaign. Sections~\ref{sec:atoll} and \ref{sec:Z} focus on atoll and Z sources, respectively. Section~\ref{sec:peculiar} presents the most peculiar objects. Section~\ref{sec:discussion} concludes our review with a discussion of lessons learned and open~questions.

\section{Atoll~Sources}\label{sec:atoll}
The first NS-LMXB observed by IXPE was the atoll source GS~1826$-$238 \citep{capitanio2023}, later followed by 
GX~9+9 \citep{chatterjee2023,ursini2023},
4U~1820$-$303 \citep{dimarco2023_4U1820},
and
4U~1624$-$49 \citep{saade2024}. All of them were observed in the soft state. We report in Table~\ref{tab:poldeg} the integrated 2--8 keV PD, as~well as the maximum PD measured in a specific energy band. The~2--8 keV PD is usually low (below 1--2\%, with~the exception of the dipping source 4U~1624$-$49); however, there is a trend of PD increasing with energy, reaching 10\% in 4U~1820$-$303. 

\begin{table}[H] 
\caption{The polarization degree of the atoll (top part of the table), Z sources (middle part of the table), and~peculiar sources (bottom part of the table).\label{tab:poldeg}}
\begin{tabularx}{\textwidth}{LCCC}
\toprule
\textbf{Source} &  \textbf{Integrated PD} \textbf{(2--8~keV)} & \textbf{max PD (Band)} & \textbf{Reference} \\
\midrule
GS 1826$-$238 (soft)  & $<1.3\%$ & -  & \cite{capitanio2023} \\
GX 9+9 (soft) & $1.4\% \pm 0.3\%$ & $2.2\% \pm 0.5\%$ (4--8 keV) & \cite{ursini2023} \\
4U 1820$-$303 (soft) & $<1.3\%$ & $10.3\%\pm2.4\%$ (7--8 keV) & \cite{dimarco2023_4U1820}  \\
4U 1624$-$49 (soft) & $3.1\% \pm 0.7\%$ & $6\% \pm 2\%$ (6--8 keV) & \cite{saade2024}  \\
\midrule
Cyg X-2 (NB) & $1.8\% \pm 0.3\%$ & $2.8\% \pm 0.6\%$ (4--8 keV) & \cite{farinelli2023} \\
XTE J1701$-$462 (HB) & $4.6\% \pm 0.4\%$ & PD $ \sim$ const.~\textsuperscript{1} & \cite{cocchi2023} \\
XTE J1701$-$462 (NB/FB) & $<1.5\%$ & - & \cite{cocchi2023} \\
GX 5$-$1 (HB) & $4.3\% \pm 0.3\%$ &  $5.4\% \pm 0.7\%$ (5--8 keV) & \cite{Fabiani24} \\
GX 5$-$1 (NB/FB) & $2.0\% \pm 0.3\%$ &  $2.6\% \pm 0.7\%$ (5--8 keV) & \cite{Fabiani24} \\
Sco X-1 (SA/FB) & $1.0\% \pm 0.2\%$ & PD $ \sim$ const. & \cite{lamonaca2024} \\
\midrule
Cir X-1 &  $1.6\% \pm 0.3\%$ & PD $ \sim$ const.~\textsuperscript{2} & \cite{rankin2024} \\
GX 13+1 & $1.4\% \pm 0.3\%$ & strongly variable~\textsuperscript{3} & \cite{bobrikova2024} \\
\bottomrule
\end{tabularx}
\noindent{\footnotesize{\textsuperscript{1} However, a~PD of 6\% is attributed to the hard component.}}
\noindent{\footnotesize{\textsuperscript{2} The PA, however, varies with time.}}
\noindent{\footnotesize{\textsuperscript{3} The PD varies both with time and energy (see text for details).}}
\end{table}

\begin{figure}[H]
\includegraphics[width=0.95\textwidth]{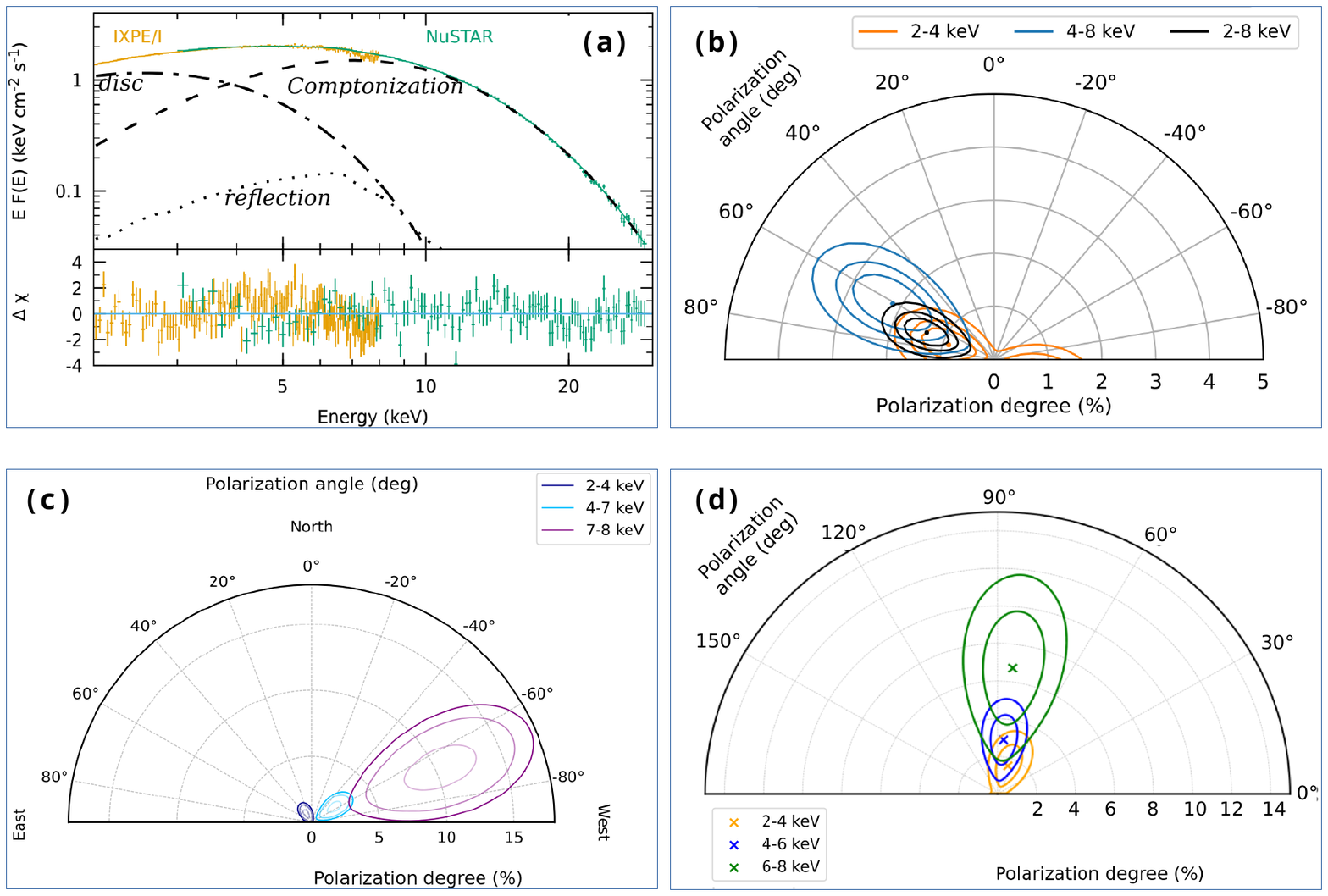}
\caption{
(\textbf{a}) IXPE and NuSTAR spectral data of GX~9+9 fitted with a model consisting of disk emission, Comptonization, and reflection (adapted from \citep{ursini2023}). 
(\textbf{b}) PD$-$PA contour plots of GX~9+9 (68, 90, and 99\% confidence levels; adapted from \citep{ursini2023}), 
(\textbf{c}) 4U~1820$-$303 (50, 95, and 99\% confidence levels; adapted from \citep{dimarco2023_4U1820}), and~
(\textbf{d}) 4U~1624$-$49 (68 and 95\% confidence levels; adapted from \citep{saade2024}). \label{fig:atolls}
}
\end{figure}

GS~1826$-$238 and GX~9+9 were observed by IXPE for 100~ks each in March and October 2022, respectively. Both were jointly observed by NICER, NuSTAR, and/or INTEGRAL, ensuring  broad-band spectral coverage \citep{capitanio2023,ursini2023}. The~two sources show similar spectral properties, the~most notable difference being the presence of reflection features: these are seen in GX~9+9, but not in GS~1826$-$238~\citep{capitanio2023,ursini2023} (see Figure~\ref{fig:atolls}a). In~GS~1826$-$238, IXPE finds only an upper limit of PD $<1.3\%$, slightly lower than the PD measured for GX~9+9~\citep{capitanio2023}. The~polarization of GX~9+9 also shows a hint of an increasing trend with energy, reaching 2--3\% in the 4--8 keV band \citep{chatterjee2023,ursini2023} (Figure~\ref{fig:atolls}b). These results can be explained by the combination of Comptonization in a BL and reflection \citep{ursini2023}. In~fact, the~BL emission is not expected to be highly polarized \citep{gnarini2022,farinelli2024}, while the reflected radiation can provide a significant contribution to polarization at high energies \citep{lapidus&sunyaev1985,schnittman&krolik2009}. The~higher polarization of GX~9+9 compared with GS~1826$-$238 can, thus, be related to the stronger reflection component \citep{capitanio2023_review}. 

In all sources, a low polarization is generally found below 3--4~keV, where the thermal emission dominates. So far, 4U~1820$-$303, a~bright ultracompact NS-LMXB, offers the most dramatic example of this behavior \citep{dimarco2023_4U1820}. This source was observed by IXPE twice, in~October 2022 (16 ks) and April 2023 (86 ks), in~the same spectral state. The~PD strongly increases in energy, from~$<1\%$ below 4~keV up to the surprisingly high value of 10\% in the 7--8~keV band \citep{dimarco2023_4U1820}. The~measurements of PA indicate that the soft and hard components are polarized orthogonally to each other \citep{dimarco2023_4U1820} (Figure~\ref{fig:atolls}c). 4U~1624$-$49 is another intriguing case. It was observed by IXPE in August 2023 for 200~ks, and it shows a PD of 3\% in the 2--8~keV band, with~an indication for an increase with energy, up~to 6\% in the 6--8~keV band \citep{saade2024} (Figure~\ref{fig:atolls}d). 
4U~1624$-$49 is a dipping source seen at a high inclination, but~even in this case, a~BL alone cannot explain the high PD \citep{gnarini2022,farinelli2024}. Possible scenarios to explain the high polarization include the presence of significant reflection and/or a radially extended X-ray emitting region, such as a corona \citep{saade2024} or scattering in a wind~\cite{tomaru2024}. 
Another possible scenario is that Comptonization takes place in an outfowing plasma with a mildly relativistic velocity, which can produce a PD up to $\sim 10\%$ \cite{poutanen2023}. This could, in principle, explain the high PD of 4U~1820$-$303, however, the question of why the PD is high only in the 7--8 keV band is still difficult to solve~\cite{dimarco2023_4U1820}. As~for 4U~1624$-$49, the~spectropolarimetric fit favors a model including soft disk emission, a~hard Comptonized component from a BL/SL, and~disk reflection \cite{gnarini2024}.

\unskip

\section{Z~Sources}\label{sec:Z}

The first Z sources observed by IXPE included three persistent NS-LMXBs, namely
Cygnus~X-2 \citep{farinelli2023},
GX~5$-$1 \citep{Fabiani24}, 
Scorpius~X-1 \citep{lamonaca2024}, 
and the transient XTE~J1701$-$462 \citep{cocchi2023,jayasurya2023}. The~measured PDs are reported in Table~\ref{tab:poldeg}. 

\begin{figure}
\includegraphics[width=0.9\textwidth]{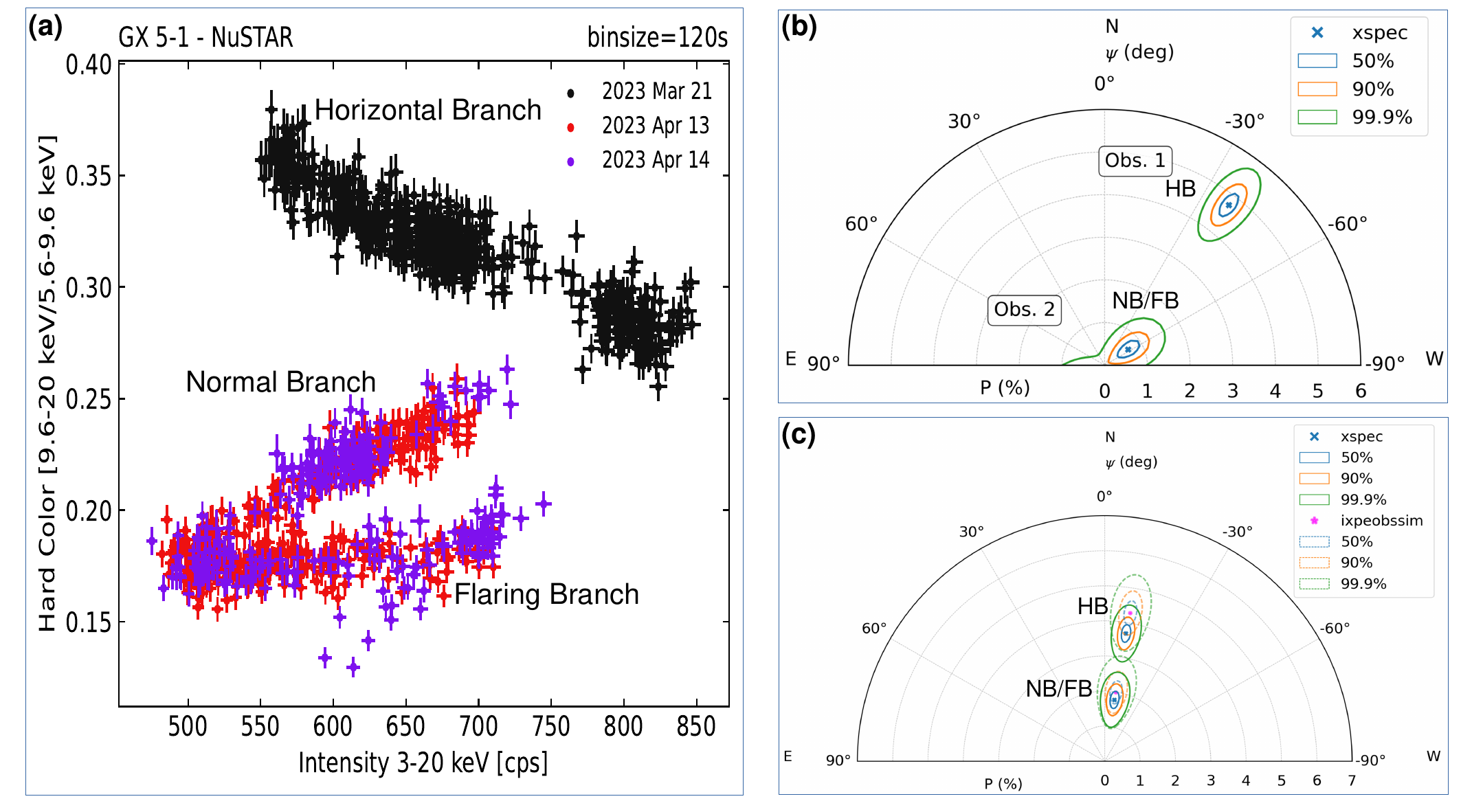}
\caption{(\textbf{a}) Typical Z track in the hardness$-$intensity diagram of GX~5$-$1, based on NuSTAR data simultaneous to IXPE observations (adapted from \citep{Fabiani24}).  (\textbf{b}) PD$-$PA (denoted here as P and $\Psi$) contour plots in the 2--8 keV band of XTE~J1701$-$462, showing a correlation with the position along the Z track (adapted from \citep{cocchi2023}). (\textbf{c}) PD-PA contour plots of GX~5$-$1 in the 2--8 keV band, showing a similar behavior (adapted from \citep{Fabiani24}). In the latter plot, solid and dashed lines indicate contours obtained with \textsc{xspec} and \textsc{ixpeobssim} \cite{baldini2022}, respectively. \label{fig:Z}}
\end{figure}

IXPE observations enabled polarization measurements along the whole Z track, typically consisting of a Horizontal Branch (HB), a~Normal Branch (NB) and a Flaring Branch (FB); HB is the hardest spectral state, while FB is the softest (see Figure~\ref{fig:Z}a).
IXPE revealed that the polarization is correlated with the position
along the Z track: PD is high (up to 4--5\%) for HB \citep{cocchi2023,Fabiani24}, and~low (1--2\%) for NB and FB \citep{farinelli2023,cocchi2023,Fabiani24,lamonaca2024}. This correlation was first observed in XTE~J1701$-$462, a~transient NS-LMXB that was targeted by IXPE twice for 50~ks in September/October 2022, three weeks after the onset of an outburst. During~the first pointing, the~source was in HB and exhibited a PD of 4.6\% in the 2--8~keV band \citep{jayasurya2023,cocchi2023}. Ten days later, the~source was observed in NB and the PD in the 2--8~keV band had dropped to $<1.5\%$ \citep{cocchi2023} (Figure \ref{fig:Z}b). A~similar behavior was observed also in GX~5$-$1, which was targeted by IXPE twice for 50~ks, in~March and April 2023. As~in the case of XTE~J1701$-$462, the~energy-integrated PD is significantly higher when the source is in the HB (Figure \ref{fig:Z}c). GX~5$-$1 also displays a puzzling behavior for PA. In~fact, the~PA in the 2--3 keV band shows a $\sim20\degr$ difference compared with the PA at higher energies \citep{Fabiani24}. 
Because in systems hosting a NS radiative processes occur at distances of at least 3 Schwarzschild radii, energy-dependent PA behavior would be difficult to attribute to  general and special relativistic effects (which actually become effective at smaller distances), but~must originate in geometrical configurations where components are not strictly orthogonal each other. Disk precession or warping with respect to the NS spin axis is a very intriguing explanation, albeit reproducing such effects via detailed Monte Carlo simulation is a challenging~task.

IXPE measured a lower polarization degree (1--2\%) in Cyg~X-2 and Sco~X-1, both observed in the soft branches only (NB and FB, respectively). These two sources are among the few NS-LMXBs known to host resolved radio jets \citep{fomalont2001,spencer2013}. This enables the comparison between the X-ray PA and the direction of the radio jet, which is assumed to be the same as the symmetry axis of the system. Cyg~X-2 was targeted by IXPE in April/May 2022 for 76~ks. Remarkably, in~this source, PA is consistent with the orientation of the radio jet, which is expected to be parallel to the disk axis \citep{farinelli2023}. The~PA is also consistent with the previous measurement by OSO-8~\cite{long1980}, but~with greater significance. 
Additionally, the~spectropolarimetric fit provided a hint of a $90^{\circ}$ swap between the PA of the soft and hard components, the latter being aligned with the radio jet. These results were interpreted in the framework of a spectropolarimetric model, consisting of a standard disk for the soft component, and~Comptonization in a BL/SL for the hard component~\cite{farinelli2023}. The~PA of the soft component is indeed consistent with direct disk emission, while the PA of the hard component indicates that Comptonization takes place in a vertically extended BL/SL rather than in a region coplanar with the disk~\cite{farinelli2023}. Disk reflection can also be significantly polarized; unfortunately, this component was unconstrained owing to the limited statistic~\cite{farinelli2023}. However, as Comptonization in a BL/SL is expected to produce a low PD (not exceeding 1\%), it is reasonable to infer that disk reflection contributes to the net polarization signal in the IXPE band~\cite{farinelli2024}. 

Sco~X-1, the~brightest persistent source in the X-ray sky, was observed by IXPE in August 2023 for 24~ks, while it was mostly in the Soft Apex (SA) that connects NB to FB. The~previous measurement with PolarLight produced a PD of 4\% and PA parallel to the radio jet, at~least in the high flux state, and~a non-detection during a low flux state \citep{long2022}. IXPE instead measured a PD of 1\%, with~no evidence of a significant trend with energy, and~a PA rotated by 40\degr--50\degr\ with respect to the jet \citep{lamonaca2024}. 
The most likely explanation for such PA misalignment is the presence of a preceding disk \citep{lamonaca2024}. Following the trend observed in XTE~J1701$-$462 and GX~5$-$1, we can reasonably infer that the higher PD previously measured by PolarLight was due to the source being predominantly in HB, albeit the relatively narrow band of PolarLight did not allow for
branch identification \citep{long2022}.

\section{Peculiar~Sources}\label{sec:peculiar}
\begin{figure}[H]
\includegraphics[width=\textwidth]{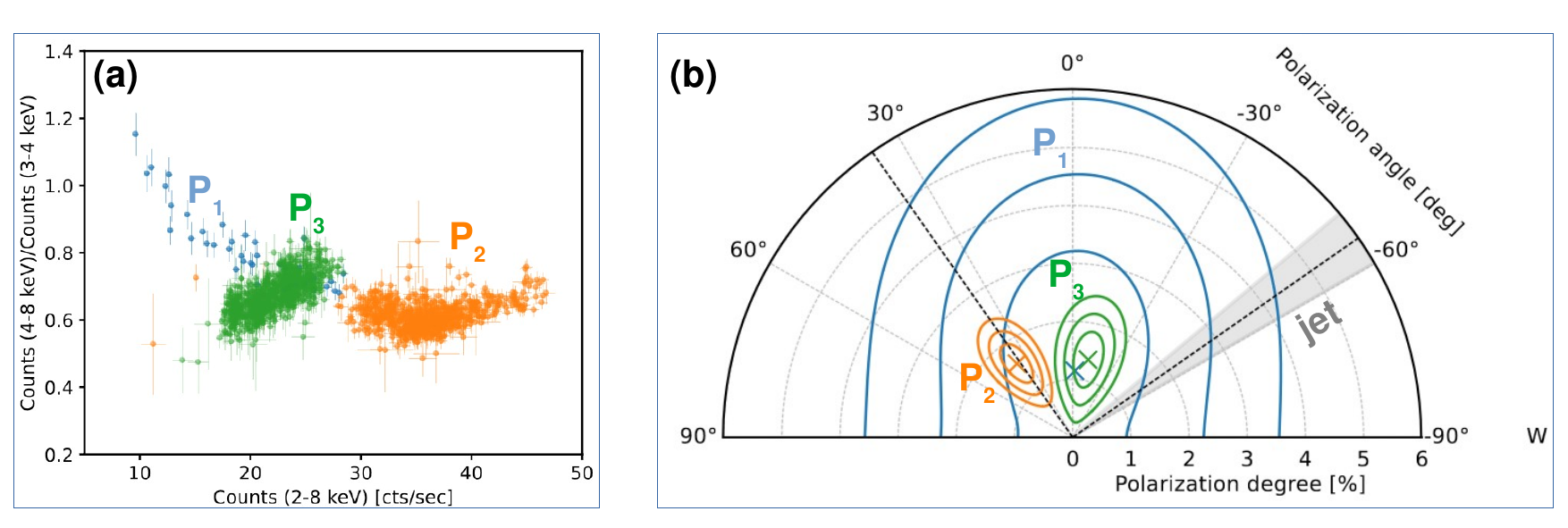}
\caption{
(\textbf{a}) Hardness$-$intensity diagram of Cir~X-1 during the IXPE observations, highlighting the three different phase intervals (adapted from \citep{rankin2024}). 
(\textbf{b}) PD$-$PA contour plots of  Cir~X-1 (68, 95, and 99\% confidence levels; adapted from \citep{rankin2024}). \label{fig:cirx1}
}
\end{figure}

IXPE observed two peculiar objects that exhibit properties of both atoll and Z sources, namely 
Cir~X-1 \citep{rankin2024}
and
GX~13+1 \citep{bobrikova2024}. 
Cir~X-1 was targeted during August 2023 in two pointings separated by six days, to~cover two different parts of the orbit, for~a total exposure time of 263~ks. GX~13+1 was observed continuously for 100 ks in October 2023.
One of the most intriguing results is the observation of a significant variation of PA with time \citep{rankin2024,bobrikova2024}. 
Along the orbit of Cir~X-1, 
the polarization is well constrained in the high-flux, soft phases (P2 and P3 in Figure~\ref{fig:cirx1}a). The~PA is found to change by $\sim$50\degr\ between different phase intervals, while PD is 1--2\% and is consistent with the remaining constant \citep{rankin2024} (Figure~\ref{fig:cirx1}b).
As for GX~13+1,~PA is found to rotate by $\sim$70\degr\ during the two-day IXPE exposure; in this case,~PD also varies in time, ranging from non-detectable during a dip in flux seen in the IXPE light curve, and~up to 5\% in the phase after the dip \citep{bobrikova2024}. Differently from Cir~X-1, these variations in polarization properties are not related to clearly separate spectral states \citep{bobrikova2024}.
These results challenge theoretical models, however, they broadly indicate the coexistence of two variable components with misaligned symmetry axes \citep{rankin2024,bobrikova2024}. 
\unskip

\section{Summary}\label{sec:discussion}
The IXPE satellite enabled a big step forward in our knowledge of accretion processes in weakly magnetized NS-LMXBs.
In particular, it revealed that 
 the hard emission component dominates the polarization signal, while the soft emission has a lower polarization \citep{farinelli2023,cocchi2023,ursini2023,dimarco2023_4U1820,lamonaca2024}. 
This trend is apparent in both atoll and Z sources. 
The spectropolarimetric data are generally well described by the combination of soft, thermal emission from the accretion disk and Comptonization in a BL or SL, with~a possible contribution from reflected radiation \citep{capitanio2023,farinelli2023,ursini2023,Fabiani24}. This geometry naturally produces a PA parallel to the jet~\citep{farinelli2023}. On~the other hand, the~unexpectedly high PD measured in some sources is not easy to explain; scattering in an extended corona, or~perhaps in a wind, seen at high inclination, cannot be ruled out~\cite{dimarco2023_4U1820,saade2024}. Another puzzling result is the energy and time dependence of PA observed in some sources, challenging simple models based on axially symmetric geometries \citep{Fabiani24,lamonaca2024,rankin2024,bobrikova2024}. A~misalignment between the NS spin and the orbital axis has been suggested as a possible scenario \citep{rankin2024,bobrikova2024}. In~any case, the~IXPE polarimetric measurements provide constraints to the accretion geometry and represent a powerful test for theoretical predictions~\cite{gnarini2022,farinelli2024,tomaru2024}, paving the way for further~developments.

\vspace{6pt} 




\authorcontributions{Writing---original draft preparation, F.U.; writing---review and editing, A.G., F.C., A.B., M.C., A.D.M., S.F., R.F., F.L.M., J.R., M.L.S. and J.P. All authors have read and agreed to the published version of the manuscript.}

\funding{F.U., A.G., F.C., A.D.M., and F.L.M. acknowledge financial support by the Italian Space Agency (Agenzia Spaziale Italiana, ASI) through contract ASI-INAF-2022-19-HH.0. F.C. acknowledges support by the Istituto Nazionale di Astrofisica (INAF) grant 1.05.23.05.06: ``Spin and Geometry in accreting X-ray binaries: The first multi frequency spectro-polarimetric campaign''. 
A.B. was supported by the Finnish Cultural Foundation grant 00240328.  
A.B. and J.P. thank the Academy of Finland grant 333112 for support. 
A.D.M., S.F., and F.L.M. aknowledge partial financial support by MAECI with grant CN24GR08 “GRBAXP: Guangxi-Rome Bilateral Agreement for X-ray Polarimetry in Astrophysics”.}

\dataavailability{No new data were created or analyzed in this study. Data sharing is not applicable to this article.}

\acknowledgments{We thank the referees for constructive comments that improved the~manuscript.}

\conflictsofinterest{The authors declare no conflicts of interest. The funders had no role in the design of the study; in the collection, analyses, or interpretation of data; in the writing of the manuscript; or in the decision to publish the results.}



\abbreviations{Abbreviations}{
The following abbreviations are used in this manuscript:\\

\noindent 
\begin{tabular}{@{}ll}
NS & Neutron star\\
LMXB & Low-mass X-ray binary\\
PD & Polarization degree\\
PA & Polarization angle
\end{tabular}
}

\begin{adjustwidth}{-\extralength}{0cm}

\reftitle{References}



\PublishersNote{}
\end{adjustwidth}
\end{document}